\documentstyle[twoside]{article}
\begin{document}
\hspace{7.5cm} Preprint Bu-He 93/5
\vspace{1.5cm}
\large
\bf
\begin{center}
Phase Transition Study of Superconducting Microstructures
\end{center}
\normalsize
\vspace{2cm}
C. Berger, A. Gabutti, S. Janos\\
{\it Laboratory for High
 Energy Physics, University of Berne, CH-3012 Berne, Switzerland\\}
P. Carelli\\
{\it Dipartimento di Energetica, Universit\a`a di L'Aquila,
Italy\\}
M.G. Castellano, R. Leoni, D.~Peschiaroli\\
{\it IESS CNR, Via Cineto Romano 42, I-00156 Rome, Italy}
\rm
\begin{center}
\section*{Abstract}
\end{center}
The presented results are part of a feasibility study of superheated
superconducting microstructure detectors. The microstructures (dots) were
fabricated using
thin film patterning techniques with diameters ranging from $50\mu$m up to
$500\mu$m and thickness of $1\mu$m. We used arrays and single dots to study the
dynamics of the superheating and supercooling phase transitions in a magnetic
field parallel to the dot surface. The phase transitions were produced by
either varying the applied magnetic field strength at a constant temperature
or changing the bath temperature at a constant field. Preliminary results
on the dynamics of the phase transitions of arrays
and single indium dots will be reported.

\section{INTRODUCTION}
Metastable superconducting devices, either made of a suspension of granules
or consisting of thin film arrays, are presently under investigation for
different applications like dark matter detection \cite{Pretzl}
or high spatial resolution x-ray imaging devices \cite{Hueber}.
Experiments with single granules have shown that the strength of the
superheating and supercooling fields depends on the crystalline structure
and on the orientation of the granule with respect to the applied magnetic
field \cite{granor}. The measured superheating
field distributions of Superheated Superconducting Granule (SSG) detectors
have standard deviations of the order of 15-20\%.
An order of magnitude improvement on the
phase transition spread of a collection of granules can be obtained if
ordered arrays made of spherically melted indium pads (PASS detector)
are used \cite{Legros}.

The metastability of thin films has been studied in the past by several
authors using strips or square samples \cite{metastinfilm}. In particular,
J. Blot et al. \cite{Blot} investigated the superheating and
supercooling transition of thin indium squares in
perpendicular and parallel magnetic fields at temperatures very close to
$T_{c}$  with susceptibility measurements.
In the earlier experiments, the phase transition of the whole array was
measured
without extracting informations on the behaviour of single elements.
To investigate the dependence of the metastable phase transitions
on the granularity and on the geometrical order of the
superconducting detector elements, we are studying arrays made of aluminum
and indium thin disks (dots) exposed to  perpendicular and parallel
magnetic field.
The readout configurations used in our experiments allow us to detect
the phase transition of each element inside the array.
Recent measurements of indium dots in perpendicular magnetic field, with the
readout coil integrated directly on the substrate, are reported
in \mbox{Ref.\cite{Leoni}}. Preliminary results
on the dynamics of the phase transitions of arrays
and single indium dots in a parallel field will be reported in this paper.

\section{EXPERIMENTAL PROCEDURE}
We measured indium dots of cylindrical shape with a thickness
of 1$\mu$m and diameters between  50$\mu$m and 500$\mu$m.
The films were fabricated at the IESS-CNR in Rome on silicon substrates
using electron-beam evaporation and patterned with photolithography
and lift-off technique.
Scanning electron microscope analysis have shown a
granular structure of the indium film with a typical grain size of
2-3$\mu$m. The geometrical definition of the edges of the dots was
dominated by the film granularity.
The experiments were performed in a pumped $^{4}$He cryostat. The
temperature was determined by measuring both the vapor pressure over the
$^{4}$He bath and the value of a calibrated Allan-Bradley 105$\Omega$ resistor.
The investigated range of temperature was from 1.4K up to 2.1K.
The temperature regulation was done by adjusting the pumping
power with a regulating valve and by heating the $^4$He bath.
The temperature stability was better than 10mK.
The samples were mounted at the end of a top-loading insert. The magnetic
field was produced by a superconducting Helmholtz coil placed inside
the cryostat. The homogeneity of the field was calculated to be
within $10^{-3}$ in the volume surrounding the sample.

We performed measurements applying the magnetic field at different angles
with respect to the dot surface. The angles were varied by turning the
top-loading insert and measured with a laser beam incident on a
mirror placed on the axis of the insert. The angular resolution of the setup
was better than 0.03$^{\circ}$.
To measure the change in flux due to the transition of a single dot, a one or
a three layer pickup coil was wound directly around the
Si substrate as it is shown in \mbox{Fig. 1$a$}.
The coils were made of 63-84 turns of $25\mu$m copper wire and wound under
a microscope with the help of a micromanipulator. The precision of the
wire alignement was better than $10\mu$m.
The distance between the dot surface and the windings of the coils was
measured to be  about $0.5\mu$m. In a few samples, only some rows of
dots were covered by the pickup coil as it is shown in
\mbox{Fig. 1$b$}. In  this configuration the measured change in
flux of a dot covered by the pickup coil was higher
compared to the one of a dot outside the coil. This allowed us to identify the
row of the flipped dot. The coil was coupled to a current
sensitive amplifier with a risetime of $\simeq 80$ns and a gain of 2000.
\section{MEASUREMENTS}
Thin superconducting dots in a parallel field have superheating phase
transitions (flip) characterized by a fast transition time, as it is shown
in \mbox{Fig. 2$a$} for a 300$\mu$m wide and 1$\mu$m thick
In-dot at 1.4K.  The speed of the transition is comparable
to the measurements with superconducting granules \cite{Miha}.
When the sample is slightly tilted with respect to the magnetic field, the
flip becomes broader, as it is shown in \mbox{Fig. 2$b$}.
The fragmentation of the signal is an indication that the nucleation of the
superheating transition does not happen at once but rather in steps.
This is due to the increased effective field strength at the upper and lower
edges of the dot when the sample is tilted \cite{Hubbook}.
As a result the field penetration starts locally.
The angle at which the flip signal starts to be broad depends on the diameter
of the dot but not on the structure of the superconducting film.
Typical angles are  $\pm 2.5^{\circ}$ for a 50$\mu$m wide
dot and  $\pm 0.25^{\circ}$ for a 300$\mu$m sample. The speed of the
supercooling
transition (flop) was measured to be $\sim 1\mu$s for a 300$\mu$m wide dot.
It does not significantly depend on the
sample orientation in the range of angles from $0^{\circ}$ up to
$\pm 6^{\circ}$.

The distribution of the superheating and supercooling fields of the arrays
was measured by cycling the magnetic field at a constant temperature.
In each cycle the field was changed with a constant ramping speed of
40 Gauss/s from zero up to 400 Gauss and then lowered to the zero value.
We were able to observe the superheating and supercooling transition of
each dot in the array.
Typical flip and flop signals recorded with the sequence
trigger function of a digital oscilloscope (LeCroy 9450) are shown in
\mbox{Fig. 3$a$} for a 5x5 array of 300$\mu$m dots
in a parallel field at 1.4K.
The measurements were done using the
position sensitive coil readout as shown in \mbox{Fig. 1$b$}.
The area filling factor of the sample was 28\% and the dots can be
considered as isolate elements.
In each cycle the same sequence of signals was identified.
In agreement with previous measurements on collections of superconducting
granules \cite{Larrea}, it turned out that the superheating and
supercooling fields are two independent properties of each dot.
The penetration of the magnetic field in dots with low superheating
field is probably induced by geometrical defects which locally enhance the
strength of the applied field.
The time evolution of an early and a late flip signal within a cycle
is shown in \mbox{Fig. 3$b$}. The transition time of the early
flipping dots is longer because the penetration of the field starts
locally. Dots with higher superheating fields have less surface defects and
the field penetration is more symmetric and faster.
The amplitude of the flip signals is due to the position of the dots with
respect to the pickup coil and to the speed of the superheating phase
transition. The nucleation of the supercooling transition is not influenced
by such magnetostatic effects. The transition speed was always
$\sim$1$\mu$s and the amplitude modulation is only due to the dot position
with respect to the pickup coil.

The measured superheating and supercooling distributions of a 20x20 array
made of 100$\mu$m indium dots at the temperature of 1.4K are shown
in \mbox{Fig. 4}.  The area filling factor of
the array was 65\% and all the dots were covered by the
pickup coil. To increase the statistics, the measurements were done in 500
consecutive cycles. The spread of the distributions was evaluated with a
gaussian fit of the data.
Standard deviations of 2\% were found in both superheating and supercooling
field distributions.
The observed spread is an order of magnitude lower than the typical values
 ($\sigma$=15-20\%) measured in collections of granules \cite{Pretzl}.
This could be related to the polycrystalline structure of the indium film.

To investigate  magnetic interactions among dots, we performed measurements
on a 10x18 array made of 50$\mu$m dots with an area filling factor of 68\%.
The measurements were done with the array parallel and tilted by 1.5$^{\circ}$
with respect to the applied magnetic field.
In \mbox{Fig. 5} the two sequences of flipping signals
within a cycle are shown. When the array is parallel to the field  multiple
flips were observed, especially at the end of the cycle.
Multiple flips were identified by comparing the measured change in flux to
the value associated to a single dot transition. Multiple flip signals start
to split up in single flip signals when the array is tilted by
$\ge 0.5^{\circ}$.
The time evolution of the transition signal is shown in
\mbox{Fig. 5$c$} for the same event in a parallel field and
at $1.5^{\circ}$.
The contribution of three dots to the flipping signal is clearly
visible when the array is tilted. The same sequence of signals was found in
consecutive cycles also when the magnetic field was ramped at different speeds.
Multiple flips were not observed in arrays
with smaller area filling factor.

The occurrence of multiple flips in highly packed arrays can be
interpreted as a magnetic avalanche effect.
The density of the field lines changes locally when a dot undergoes a
phase transition. This effect
increases the strength of the magnetic field near a neighbour dot, producing
another phase transition. Such a mechanism can propagate to other
elements of the array. We measured a phase transition multiplicity up to four
in a parallel field. When the array is tilted the
magnetic coupling among dots is less strong.

\section{CONCLUSION}
The measured superheating and supercooling distributions of indium dot
arrays have spreads one order of magnitude lower than the typical values
measured in collections of granules. We are planning  a more systematic
study on dot arrays made of films with different granularities to investigate
if the smearing of the superheating and supercooling fields can be reduced
by using polycrystalline films.
We measured the occurrence of multiple flips
in highly packed arrays. The use of such a mechanism as a possible
signal amplification in bidimensional superheated superconducting
detectors is under investigation.

\section*{ACKNOWLEDGEMENTS}
The authors would like to thank G. Czapek and K. Pretzl for initiating
this work, K. Borer, M. Hess, E. Kr\"ahenb\"uhl and F. Nydegger
for technical support.
This work was supported by the Schweizerischer Nationalfonds zur
F\"orderung der wissenschaftlichen Forschung, the Bernische Stiftung
zur F\"orderung  der wissenschaftlichen Forschung an der Universit\"at Bern,
the National Research Council of Italy and the Istituto di Fisica Nucleare.

\section*{Figure captions}
\begin{enumerate}
\item $a$) Schematic sketch of the pickup coil arrangement used in the
  measurements. $b)$ Position sensitive pickup coil over a 5x5 array.
\item Time evolution of the superheating transition signal  $V(t)$
 (upper curves) and of the corresponding change in flux $\Phi (t)$
 (lower curves)
 for a 300$\mu$m wide and 1$\mu$m thick In-dot at 1.4K.
 $a$) Parallel field.  $b)$ Dot tilted by 0.8$^{\circ}$.
\item $a$) Sequence of flip (positive) and flop (negative) signals of a
 5x5 array made of 300$\mu$m dots at 1.4K. The measurements were done in
 a parallel field and two consecutive cycles are shown.
 $b$) Time evolution of the flip transitions  $V(t)$
 (two upper curves) and corresponding change in flux $\Phi (t)$
 (lower curves) for an early and a late event within a cycle.
\item $a$) Supercooling  and $b$) superheating
distributions measured in 500 consecutive cycles with a 20x20 array made
of 100$\mu$m dots. The standard deviations of the distributions are $2.05\%$
(supercooling) and $1.87\%$ (superheating).
\item $a$) Sequence of flipping signals in a parallel field and
          $b$) tilted by $1.5^{\circ}$.\\
          $c$) $V(t)$ for a flip with multiplicity of 3 in a parallel field
               (upper line) and for the same event with the array tilted
               by $1.5^{\circ}$ (lower line)
\end{enumerate}
\end{document}